\documentclass[
reprint,
superscriptaddress,
amsmath,amssymb,
aps,
]{revtex4-2}

\usepackage[pdftex]{graphicx}
\usepackage{dcolumn}
\usepackage{bm}
\usepackage{braket}
\usepackage{physics}
\usepackage{color}
\usepackage{cancel}

\newcommand{\fracpd}[2]{
	\genfrac{}{}{}{}{\partial #1}{\partial #2}    %
}

\usepackage{placeins}

\begin{document}
	
	\preprint{APS/123-QED}
	
	\title{Variational Quantum Operator Simulation}
	
	\author{Satoru Shoji}
    \email[Contact author: ]{satoru.shoji.t8@dc.tohoku.ac.jp}
    \affiliation{Department of Applied Physics, Graduate School of Engineering, Tohoku University, Sendai 980-8579, Japan}
    
	\author{Kosuke Ito}
    \affiliation{Center for Quantum Information and Quantum Biology, International Advanced Research Institute, The University of Osaka, 1-2 Machikaneyama, Toyonaka, Osaka 560-8531, Japan}
    
	\author{Yukihiro Shimizu}
    \email[Contact author: ]{shimizu@tohoku.ac.jp}
    \affiliation{Department of Applied Physics, Graduate School of Engineering, Tohoku University, Sendai 980-8579, Japan}
    
	\author{Keisuke Fujii}
    \affiliation{Center for Quantum Information and Quantum Biology, International Advanced Research Institute, The University of Osaka, 1-2 Machikaneyama, Toyonaka, Osaka 560-8531, Japan}
    \affiliation{Graduate School of Engineering Science, The University of Osaka, 1-3 Machikaneyama, Toyonaka, Osaka 560-8531, Japan}
    \affiliation{RIKEN Center for Quantum Computing (RQC),
	Hirosawa 2-1, Wako, Saitama 351-0198, Japan}
    
	\date{\today}
	
	\begin{abstract}
		Implementing time-evolution operators in shallow quantum circuits is important for quantum simulations. The standard method of Trotterization requires a large number of gates to achieve practical accuracy.
		Variational Quantum Simulation (VQS) is an algorithm that calculates the time evolution of a quantum state and can be executed with shallower circuits than Trotterization. However, the operator obtained by VQS evolves only a fixed initial state and is not the time evolution operator itself.
		In this paper, we propose Variational Quantum Operator Simulation (VQOS), a method to realize time evolution operators in shallow quantum circuits. This method is based on the variational principle for operators and does not require the implementation of the desired Trotter decomposition of the time evolution operator.
		We performed numerical simulations of the VQOS algorithm and successfully implemented the time evolution operator for closed systems in a quantum circuit that is up to 5 times shallower than the Trotterization.
		By providing a more practical way to implement time evolution operators, VQOS increases the applicability of near-term quantum computers.
	\end{abstract}
	
	\maketitle
	
	\section{Introduction}
	
	Quantum computers are expected to offer advantages over classical computers for certain computational tasks. In particular, Hamiltonian simulation and eigenvalue problems play a central role in quantum science, and quantum algorithms for these tasks have attracted significant attention~\cite{Feynman1982, seth1996}. While classical algorithms and approximation methods are available for many physically relevant systems, quantum approaches may provide favorable scaling or practical advantages in regimes where classical simulations become challenging. Rather than claiming asymptotic quantum advantage, our focus is on reducing circuit depth and avoiding oracle access in near-term settings.
	In particular, the implementation of time-evolution operators as quantum circuits plays a crucial role not only in the simulations of quantum dynamics but also as a key component of quantum phase estimation~\cite{kitaev1995} and quantum eigenvalue transformation of unitaries~\cite{QETU}. By using the Trotterization~\cite{Trotter, Hatano2005}, the time evolution operators can be approximated as a quantum circuit. This is one of the simplest approaches, but it requires a very large number of gates, and it is not realistic to implement long-time evolution operators with sufficient accuracy before full fault tolerance~\cite{kim2023evidence, EFTQC}.
    Regarding the deep gate problem, a method of classical pre-processing involving the optimization of the Hamiltonian simulation has been proposed\cite{Mansuroglu_2023, PhysRevResearch.5.023146}.
	
	Regarding algorithms for near-term quantum devices, one approach is Quantum Assisted Quantum Compiling (QAQC)~\cite{QAQC}. This method compiles the time evolution operators approximately into a variational quantum circuit via the minimization of the cost function given by the distance from the true time evolution operator. However, this method requires oracle access to the target time evolution operator as a quantum circuit, which is a difficult problem in itself. Furthermore, the number of circuits to be evaluated to obtain a compiled circuit cannot be known in advance, and optimization may fail due to local minima or a barren plateau~\cite{bp2018} (exponentially vanishing gradient) in the cost function.
	Local variational quantum compilation (LVQC)~\cite{LVQC, LSVQC}, a variant of QAQC, uses a Lieb-Robinson bound and causal cones to reduce the effective number of qubits, and the parameters of the variational quantum circuit are optimized by a classical simulator, which requires a large classical computation.
	Therefore, for near-term applications, it is essential to develop an algorithm that approximates the time evolution operator using a shallow circuit, without requiring prior preparation of the target operator in either quantum circuits or classical computations.
	
	Another approach for dynamics simulation is variational quantum simulation (VQS)~\cite{VQS, VQST, QAS, AVQDS}. The VQS simulates the time evolution of a quantum state using the variational principle. This method does not require prior preparation of the target operator in quantum circuits. However, VQS only yields the time evolution of a fixed initial state, not the time evolution operator itself. In other words, the quantum circuit obtained by VQS approximates the time evolution from a predetermined state but does not correctly time-evolve another initial state.
	Therefore, VQS cannot be used for quantum phase estimation or long-time evolution via repeated application of the time-evolution operator.
	
	To address the above issues, this paper proposes Variational Quantum Operator Simulation (VQOS), a method to variationally approximate the target time evolution operator itself without oracle access to the target time evolution operator. In this paper, we refer to the ordinary VQS as Variational Quantum State Simulation (VQSS), in contrast to VQOS.
	Similarly to VQSS, VQOS requires neither oracle access to the target time evolution operator nor classical optimization of a cost function. Hence, it is free from trainability problems such as local solutions and the barren plateau. Furthermore, the number of quantum circuits required to run VQOS can be pre-computed from the Hamiltonian and number of variational parameters.
	This is achieved by applying the variational principle of the time evolution operator to a parameterized quantum circuit (ansatz) $U(\bm\theta(t))$.
    Although the formulation is formally equivalent to applying VQSS to the Choi state, the actual implementation can be performed on a single system without entangled inputs, resulting in a simpler and shallower circuit.
	
	We assess the performance of VQOS through classical simulations of the transverse-field Heisenberg model with up to nine sites. The results demonstrate that VQOS achieves up to three orders of magnitude improvement in accuracy compared to Trotterization, given the same circuit depth. Additionally, VQOS is capable of approximating the time-evolution operator with comparable accuracy to that of Trotterization while requiring only approximately one-quarter of the circuit depth. Importantly, the degradation in accuracy due to increasing system size was found to be minimal, indicating favorable scalability of the method.
	
	The remainder of this paper is organized as follows.
	Sec.~\ref{sec:VQS} briefly summarizes the VQSS.
	Sec.~\ref{sec:VQOS} introduces the VQOS method, and Sec.~\ref{sec:imp} proposes circuit implementations for VQOS,
	Sec.~\ref{sec:sim} presents the numerical results of applying VQOS to the transverse field Heisenberg model.
	Finally, Sec.~\ref{sec:conc} summarizes the paper.
	
	\section{Brief review of the VQSS}\label{sec:VQS}
	VQS~\cite{VQS,VQST} is a quantum--classical hybrid algorithm used to approximate the time evolution of a quantum system from a fixed initial state based on the variational principle. We refer to this conventional VQS as VQSS.
	
	\subsection{Formulations}
	The target problem is to solve the time-dependent Schrödinger equation
	\begin{equation}
		\frac{d}{dt}\ket{\psi(t)} = -i H \ket{\psi(t)},
		\label{eq:Schrodinger}
	\end{equation}
	with Hamiltonian $H$ and an initial state $\ket{\phi_{\mathrm{in}}} = \ket{\psi(0)}$.
	In VQSS, an ansatz
	\begin{eqnarray}
		\ket{\phi(\theta_0(t), \bm \theta(t))}
		&= e^{i\theta_0(t)}U(\bm\theta (t)) \ket{\phi_{\mathrm{in}}},
		\label{eq:trial_function_pure}
	\end{eqnarray}
	approximates the exact time evolution $\ket{\psi(t)}$. Here, $U(\bm\theta(t))$ is a parameterized unitary gate. The parameters are represented as a real vector of length $p$. In Eq.~\eqref{eq:trial_function_pure}, the global phase $e^{i\theta_0(t)}$ is introduced to deal with the global phase mismatch pointed out in Ref.~\cite{VQST}. Although VQSS can also be applied to time-dependent Hamiltonians, we focus on the time-independent case here for simplicity.  We use McLachlan's variational principle~\cite{McLachlan1964}
	\begin{equation}
		\delta \left\| \frac{d}{dt}\ket{\phi(\theta_0(t), \bm \theta(t))} + i H \ket{\phi(\theta_0(t), \bm \theta(t))} \right\| = 0, \label{McLachlanVP}
	\end{equation}
	to find parameters $\bm\theta(t)$ that approximately satisfy Eq.~\eqref{eq:Schrodinger} within the expressibility of the ansatz. Although there are other variational principles, we use McLachlan's due to its consistency with open systems~\cite{VQST}. As a result, we obtain a formula for updating the variational parameters
	\begin{equation}
		M \dot{\bm\theta} = V \label{paramUpdate}.
	\end{equation}
	Here, each element of the real symmetric matrix $M$ and real vector $V$ can be written as follows:
	\begin{eqnarray}
		M_{j,k} &=& \mathrm{Re} \left( \frac{\partial\bra{\phi(\bm\theta)}}{\partial\theta_j} \frac{\partial\ket{\phi(\bm\theta)}}{\partial\theta_k} \right)\nonumber \\
		&&+\frac{\partial\bra{\phi(\bm\theta)}}{\partial\theta_j} \ket{\phi(\bm\theta)} \fracpd{\bra{\phi(\bm\theta)}}{\theta_k} \ket{\phi(\bm\theta)}, \label{M}
	\end{eqnarray}
	\begin{eqnarray}
		V_k &=& \mathrm{Im}\left( \frac{\partial\bra{\phi(\bm\theta)}}{\partial\theta_k}  H  \ket{\phi(\bm\theta)} \right)\nonumber \\
		&&+i\frac{\partial\bra{\phi(\bm\theta)}}{\partial\theta_k}
		\ket{\phi(\bm\theta)}\bra{\phi(\bm\theta)}  H  \ket{\phi(\bm\theta)}. \label{V}
	\end{eqnarray}
	Here, $M$ is an $p \times p$ matrix, and $V$ is a length $p$ vector. Note that because $\frac{\partial\bra{\phi(\bm\theta)}}{\partial\theta_j} \ket{\phi(\bm\theta)}$ in Eqs.~\eqref{M} and \eqref{V} is pure imaginary, $M$ and $V$ are a real matrix and vector, respectively.
	In the VQSS, we estimate $M$ and $V$ using quantum computers. Then, Eq.~\eqref{paramUpdate} is solved classically.
	The implementation of the quantum circuit to estimate $M$ and $V$ is described in detail in a later section.
	
	Using an ansatz with sufficiently high expressibility, we get variational parameters $\bm\theta(t)$ that approximate the time evolution of states. For example, for VQSS of closed systems, the approximation
	\begin{equation}
		e^{-i H t} \ket{\phi_{\mathrm{in}}} \approx U(\bm\theta(t)) \ket{\phi_{\mathrm{in}}}
	\end{equation}
	is satisfied. However, the operator $U(\bm\theta(t))$ obtained by VQSS with the initial state $\ket{\phi_{\mathrm{in}}}$ does not necessarily give the correct time evolution for another different initial state. In other words, because VQSS does not approximate the time evolution operator itself, $e^{-i H t} \approx U(\bm\theta(t))$ is generally not satisfied. Therefore, to compute the time evolution from different initial states, we must execute VQSS for each of them. Also, the $U(\bm\theta(t))$ obtained by VQSS cannot be used for algorithms such as quantum phase estimation.
	
	\subsection{Implementations}
\begin{figure}[t]
		\includegraphics[width=8cm]{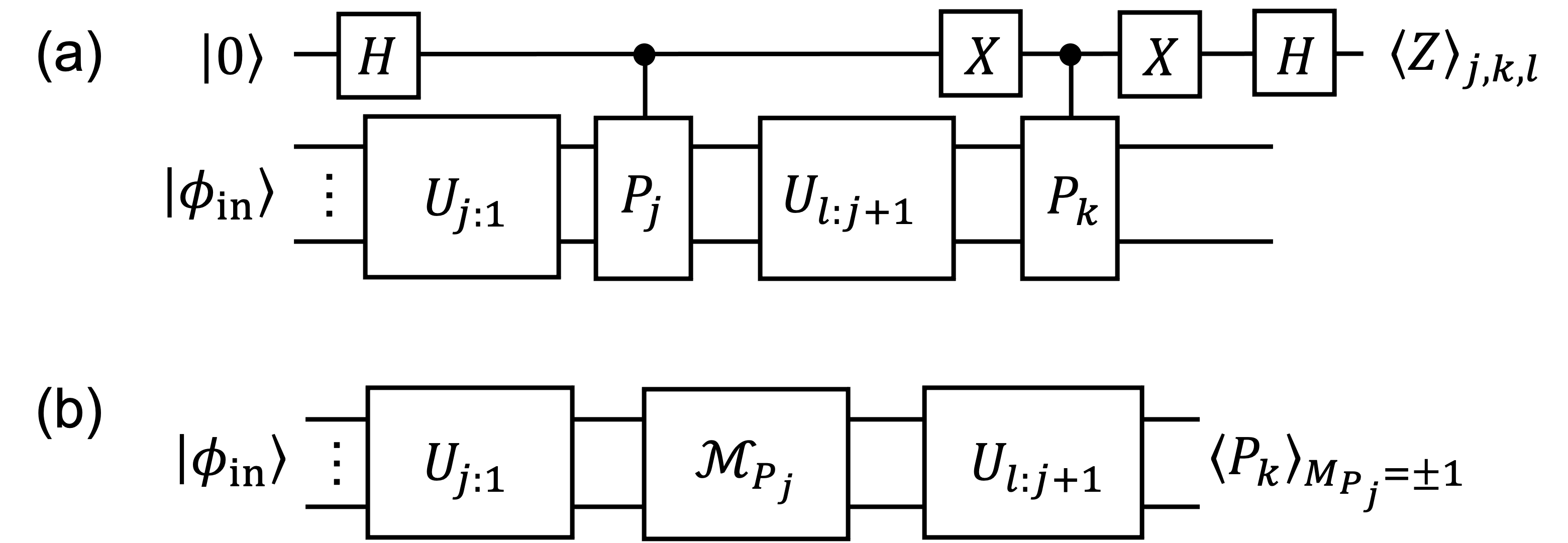}
		\caption{\label{fig:VQS_impl_circuit} Circuits to estimate Eq.~\eqref{qe:vqs_gjkl}. (a) Indirect measurement method. $g_{jkl}$ is estimated by measuring ancilla qubit $\langle Z\rangle_{j,k,l}$. (b) Direct measurement method. $\mathcal{M}_{P_j}$ is the measurement with Pauli operator $P_j$. $g_{jkl}$ is estimated as $p(M_{P_j}=+1)\langle P_k\rangle_{M_{P_j}=+1} - p(M_{P_j}=-1)\langle P_k\rangle_{M_{P_j}=-1}$. $p(M_{P_j}=\pm1)$ is the probability that the outcome of the mid-measurement is $\pm1$. $\langle P_k\rangle_{M_{P_j}=\pm1}$ is the expectation value of $P_k$ conditional on $M_{P_j}=\pm1$.}
	\end{figure}

	In VQSS, we estimate Eqs.~\eqref{M} and \eqref{V} using quantum computation. Suppose that the ansatz 
    \begin{equation}
        \ket{\phi(\bm\theta)} = R_p(\theta_p) \cdots R_1(\theta_1)\ket{\phi_\mathrm{in}},\label{eq:ansatz_def}
    \end{equation}
    is represented by the product of the rotation gates $R_j(\theta_j) = \exp(i G_j \theta_j /2),\label{eq:rot_gate_def}$
    defined by $p$ Pauli operators $G_j \in \{ I, X, Y, Z \}^{\otimes n}$. We also suppose that the Hamiltonian
    \begin{equation}
        H = \sum_jc_jH_j, \label{eq:hamiltonian_linear_combination}
    \end{equation}
    is written as a linear combination of Pauli operators $H_j \in \{ I, X, Y, Z \}^{\otimes n}$ with real coefficients $c_j$. Then, Eqs.~\eqref{M} and \eqref{V} can be written as
	\begin{align}
		M_{jk} &= \frac{1}{4}\mathrm{Re}\bra{\phi_\mathrm{in}} U_{j-1:1}^\dag G_j U_{k-1:j}^\dag G_k U_{k-1:1} \ket{\phi_\mathrm{in}}\nonumber \\
		&\hspace{1em} - \frac{1}{4}\bra{\phi_\mathrm{in}} U_{j-1:1}^\dag G_j U_{j-1:1} \ket{\phi_\mathrm{in}}\nonumber\\
		&\hspace{2em}\times \bra{\phi_\mathrm{in}} U_{k-1:1}^\dag G_j U_{k-1:1} \ket{\phi_\mathrm{in}},\label{eq:M_imp}
	\end{align}
	\begin{align}
		V_j &= \frac{1}{2}\sum_k c_k \Re \bra{\phi_\mathrm{in}} U_{j-1:1}^\dag G_j U_{p:j}^\dag H_k U_{p:1} \ket{\phi_\mathrm{in}} \nonumber\\
		&\hspace{1em} -\frac{1}{2} \bra{\phi_\mathrm{in}} U_{j-1:1}^\dag G_j U_{j-1:1} \ket{\phi_\mathrm{in}}\nonumber\\
		&\hspace{2em} \times \sum_k c_k \bra{\phi_\mathrm{in}} H_k \ket{\phi_\mathrm{in}},\label{eq:V_imp}
	\end{align}
	where $R_k(\theta_k)\cdots R_j(\theta_j)$ is written as $U_{k:j}$.
	The first terms of Eqs.~\eqref{eq:M_imp} and \eqref{eq:V_imp} can be expressed as a linear combination of
	\begin{equation}
		g_{jkl} =  \mathrm{Re} \bra{\phi_\mathrm{in}} U_{j-1:1}^\dag P_j U_{l:j}^\dag P_k U_{l:1} \ket{\phi_\mathrm{in}}.\label{qe:vqs_gjkl}
	\end{equation}
	This quantity can be estimated either by an indirect measurement method via a Hadamard test with two controlled gates ~\cite{VQST} (Fig.~\ref{fig:VQS_impl_circuit}(a)) or by a direct measurement method ~\cite{direct_measure} (Fig.~\ref{fig:VQS_impl_circuit}(b)). The second terms of Eqs.~\eqref{eq:M_imp} and \eqref{eq:V_imp} can be evaluated by estimating the expectation values $\bra{\phi_\mathrm{in}} U_{j-1:1}^\dag G_j U_{j-1:1} \ket{\phi_\mathrm{in}}$, and $\bra{\phi_\mathrm{in}} H_k \ket{\phi_\mathrm{in}}$.
	
	\section{Variational Quantum Operator Simulation}\label{sec:VQOS}
	In this section, we show that the time evolution operator can be approximated by VQOS.
	The objective of VQOS is to find a parameter $\bm\theta(t)$ of the circuit $\tilde{U}(\theta_0, \bm\theta(t)) = e^{i\theta_0(t)} U(\bm\theta(t))$ that approximates
	\begin{equation}
		e^{-i H t} \ket{\phi_{\mathrm{in}}} \approx \tilde{U}(\theta_0, \bm\theta(t)) \ket{\phi_{\mathrm{in}}},\label{eq:VQOS_for_any}
	\end{equation}
    for any initial state $\ket{\phi_{\mathrm{in}}}$. Here we consider a $d$-dimensional system.
	In Eq.~\eqref{eq:VQOS_for_any}, the global phase factor $e^{i\theta_0(t)}$ is introduced, as in the case of VQSS, to address the mismatch in the global phase. Eq.~\eqref{eq:VQOS_for_any} can be expressed as follows:
	\begin{equation}
		\delta \left\| \frac{d}{dt} \tilde{U}(\theta_0, \bm\theta(t)) + i H  \tilde{U}(\theta_0, \bm\theta(t)) \right\|_\mathrm{F} = 0.
		\label{eq:McLachlanVP_OVQS}
	\end{equation}
	Eq.~\eqref{eq:McLachlanVP_OVQS} gives the equation of VQOS, which describes the time evolution of parameters,
	\begin{equation}
		N \dot{\bm\theta} = W, \label{eq:VQOS_pure_param}
	\end{equation}
	where
\begin{align}
N_{jk}
&= \Re\Tr\left[ \left(
\frac{\partial U({\bm\theta})}{\partial\theta_j}\right)^\dagger
\left(\frac{\partial U({\bm\theta})}
{\partial\theta_k} \right)\right] \nonumber \\
&+ \frac{1}{2^n}
\left\{
\Tr\left[\left(
\frac{\partial U({\bm\theta})}{\partial\theta_j}\right)^\dagger
U({\bm\theta})\right]
\right.\nonumber \\
&\left. \hspace*{1cm}\times
\Tr\left[\left(
\frac{\partial U({\bm\theta})}{\partial \theta_k}\right)^\dagger
U({\bm\theta})\right]\right\}, \label{eq:N_pure}\\
W_j 
&= \Im\Tr\left[ \left(\fracpd{U({\bm\theta})}{\theta_j}\right)^\dagger
HU({\bm\theta})\right]\nonumber \\
&+\frac{i}{2^n}
\Tr\left[\left(
\frac{\partial U({\bm\theta})}{\partial\theta_j}\right)^\dagger
U({\bm\theta}) \right] \Tr[H]. \label{eq:W_pure}
\end{align}
The derivation of Eqs.~\eqref{eq:VQOS_pure_param},
\eqref{eq:N_pure}, and \eqref{eq:W_pure} is given in Appendix A.
    
    The evolution by Eq.~\eqref{eq:VQOS_pure_param} is equivalent to applying VQSS of the Choi state~\cite{CHOI1975285, JAMIOLKOWSKI1972275}. More specifically, the time evolution of VQSS with Hamiltonian $ H  \otimes I$ ($I$ is the identity operator with the same size as $H$) and initial state as the maximally entangled state
	\begin{equation}
		\ket{\Phi} = \frac{1}{\sqrt{d}} \sum_{j=0}^{d-1} \ket{j}\otimes\ket{j},
	\end{equation}
	is equivalent to VQOS. This equivalence can be confirmed by setting ansatz $\ket{\phi(\bm\theta(t))}$ to $\left( \tilde{U}(\theta_0, \bm\theta(t)) \otimes I \right) \ket{\Phi}$ and the Hamiltonian $ H $ to $ H  \otimes I$ in Eq.~\eqref{McLachlanVP}:
	\begin{gather}
		\delta \left\| \frac{d}{dt}(\tilde{U}(\theta_0, \bm\theta(t))\otimes I) \ket{\Phi} + i( H \otimes I) (\tilde{U}(\theta_0, \bm\theta(t)) \otimes I) \ket{\Phi} \right\| \nonumber \\
		= \frac{1}{\sqrt{d}} \ \delta \left\| \frac{d}{dt} \tilde{U}(\theta_0, \bm\theta(t)) + i H \tilde{U}(\theta_0, \bm\theta(t)) \right\|_\mathrm{F} = 0,
		\label{eq:OVQS_equiv}
	\end{gather}
	which leads to the same variational principle used in VQOS for pure states, as shown in Eq.~\eqref{eq:McLachlanVP_OVQS}. Here, in Eq.~\eqref{eq:OVQS_equiv}, $d$ is the dimension of the system we are now interested in, and we use the relation $\bra{\Phi} (A \otimes I) \ket{\Phi} = \Tr(A)/d $ that holds for a general square matrix $A$.
    This equivalence might suggest that VQOS must be implemented by performing VQSS on the Choi state, which would require preparing a maximally entangled input state. However, as explained in Sec.~\ref{sec:imp}, this is not necessary. VQOS can be implemented on a single $d$-dimensional system (i.e., an $n$-qubit register) without entangled inputs.
	
	\section{implementation}\label{sec:imp}
    \subsection{Estimation of $N_{jk}$ and $W_j$}
	In this section, we describe the implementation of VQOS to approximate the time evolution operators $e^{-i H t}$ under an $n$-qubit Hamiltonian $ H $. In order to implement VQOS, we need to compute $N_{jk}$ and $W_j$ defined in Eqs.~\eqref{eq:N_pure} and \eqref{eq:W_pure}, respectively.
    Similarly to Eqs.~\eqref{eq:ansatz_def} and \eqref{eq:hamiltonian_linear_combination}, we suppose that the variational quantum circuit $U(\bm\theta)$ is represented by the product of the rotation gates as $U(\bm\theta) = R_p(\theta_p) \cdots R_1(\theta_1)$ where $R_j(\theta_j) = \exp(i G_j \theta_j/2)$ is defined by $p$ Pauli operators $G_j \in \{ I, X, Y, Z \}^{\otimes n}$ and the Hamiltonian $ H $ can be expressed by a linear combination of $n_H$ Pauli operators $H_k \in \{ I, X, Y, Z \}^{\otimes n}$ with real coefficients $c_k$. Under the above assumptions, these quantities are written as
	\begin{align}
		N_{jk} &= \frac{1}{4} \Tr (G_j U_{k:j+1}^\dag G_k U_{k:j+1}),
		\label{eq:N_imp}\\
		W_j &= \frac{1}{2} \sum_{k=1}^{n_H} c_k \Tr (G_j U_{p:j+1}^\dag H_k U_{p:j+1}),
		\label{eq:W_imp}
	\end{align}
    where we define $U_{k:j} := R_k(\theta_k)\cdots R_j(\theta_j)$.
	Because both $N$ and $W$ can be written by a linear combination of
	\begin{align}
		g_{jkl} := \frac{1}{2^n}\Tr (P_j U_{l:j+1}^\dag P_k U_{l:j+1}), \label{eq:gjkl}
	\end{align}
	where $P_j,P_k \in \{I,X,Y,Z\}^{\otimes n}$, $N$ and $W$ can be obtained by estimating $g_{jkl}$.
	
\begin{figure}[t]
		\includegraphics[width=8cm]{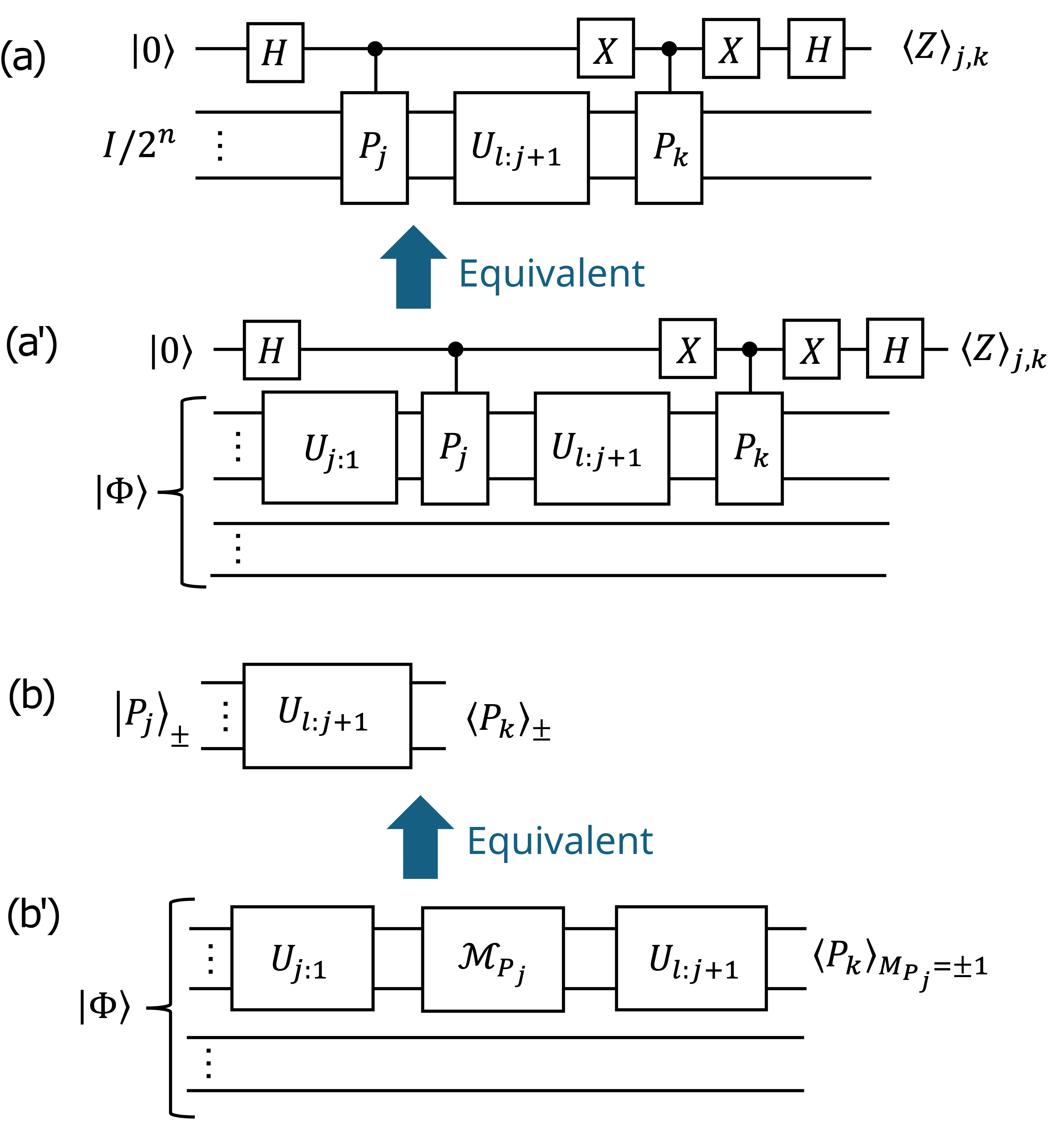}
		\caption{\label{fig:VQS_impl} Quantum circuits for estimating Eqs.~\eqref{eq:N_imp} and \eqref{eq:W_imp}. (a) A simplified, resource-reduced implementation of VQOS obtained by tracing out the idling register in (a').
        The maximally mixed input state $I/2^n$ can be realized by uniformly sampling computational basis states and averaging the measurement outcomes. 
        The expectation value of the measurement result of the ancilla qubit is $g_{jkl}$.
        (a') The circuit implementation of VQOS obtained by directly performing VQSS on the Choi state with indirect measurement.
        (b) A simplified, resource-reduced implementation of VQOS obtained by tracing out the idling register in (b') and replacing mid-circuit measurement with random initialization to eigenstates of $P_j$ belonging to the eigenvalues $+1~(-1)$.
        (b') The circuit implementation of VQOS obtained by directly performing VQSS on the Choi state with direct measurement. $\mathcal{M}_{P_j}$ represents the measurement of $P_j$. $\ev{P_k}$ is measured at the end of the circuit.}
	\end{figure}

    Although $g_{jkl}$ can be estimated by implementing VQSS of the Choi state using $n$ Bell pairs as suggested by the equivalence shown in Sec.~\ref{sec:VQOS}, VQOS admits more efficient implementations on a single $n$-qubit register without any entangled input state, as already implied by Eq.~\eqref{eq:gjkl}.
    As shown below, VQOS can be implemented by the reduced circuits in Fig.~\ref{fig:VQS_impl}(a) and Fig.~\ref{fig:VQS_impl}(b), requiring $n+1$ qubits for the indirect measurement method and $n$ qubits for the direct measurement method.
    Compared with directly implementing VQSS on the Choi state (Fig.~\ref{fig:VQS_impl}(a') and Fig.~\ref{fig:VQS_impl}(b')), these reduced circuits eliminate the entangled input and the idling register, and they are also shallower.

    First, we consider the indirect measurement method~\cite{VQS, VQST} using a single ancilla qubit.
    In the circuit implementation obtained by directly applying the indirect measurement method of VQSS to the Choi state (Fig.~\ref{fig:VQS_impl}(a')), gate operations and measurements are not performed on one qubit of each Bell pair.
    Hence, we can trace out those idling qubits to obtain the equivalent reduced circuit shown in Fig.~\ref{fig:VQS_impl}(a).
    This shows that the indirect measurement implementation only requires an $n$-qubit maximally mixed input state.
    The maximally mixed input state $I/2^n$ can be prepared by uniformly sampling the computational basis state from 
    $|0\cdots 0\rangle,\ldots,|1\cdots 1\rangle$ and then averaging the measurement results over these random inputs. This method does not require the preparation of an entangled state.
    After identifying the idling register in the Choi-state circuit, the remaining register is maximally mixed because half of the Bell pairs are maximally mixed. As the maximally mixed state is invariant under any unitary operation, $U_{j:1}(I/2^n)U_{j:1}^\dagger=I/2^n$. 
    Thus, $U_{j:1}$ does not affect the subsequent measurements and can be omitted.

    Second, the direct measurement method~\cite{direct_measure} also admits a simpler circuit implementation.
    Again, if one directly applies the direct measurement method of VQSS to the Choi state (Fig.~\ref{fig:VQS_impl}(b')), no manipulation is performed on one qubit of each Bell pair, and tracing out those qubits yields an equivalent reduced circuit.
    Furthermore, in Fig.~\ref{fig:VQS_impl}(b') the expectation value of $P_j$ is measured in the middle of the circuit.
    In VQOS, this intermediate measurement of $P_j$ can be replaced by classical randomization over a product eigenbasis of $P_j$. For each shot, we sample a bit string $b\in\{0,1\}^n$ uniformly and prepare the corresponding product eigenstate of the local Pauli operators appearing in $P_j$. The eigenvalue of $P_j$ for this state is $\lambda_j(b)=\prod_{q\in {\rm supp}(P_j)}(-1)^{b_q}$. Here, ${\rm supp}(P_j)$ denotes the set of qubits on which the Pauli string $P_j$ acts nontrivially.
    Operationally, this state is prepared by applying the appropriate basis-change gates to computational-basis states, namely $H$ for $X$, $SH$ for $Y$, and the identity for $Z$. After applying $U_{l:j+1}$, we measure $P_k$ and denote the outcome by $\mu_k=\pm1$. The sample average of $\lambda_j(b)\mu_k$ gives $g_{jkl}$. Equivalently, this procedure realizes the normalized maximally mixed states on the $\pm1$ eigenspaces of $P_j$, $\rho_\pm=(I\pm P_j)/2^n$, and estimates $(\langle P_k\rangle_+ - \langle P_k\rangle_-)/2$.
    This implementation requires only an $n$-qubit register with shallower circuit $U_{l:j+1}$ and eliminates both the entangled input state and the intermediate measurement.
    
	We also remark that the number of circuits to be evaluated in VQOS is less than that with VQSS for the same number of parameters because the second terms in Eqs.~\eqref{M} and \eqref{V} are trivially zero for VQOS using the ansatz in the form of Eq.~\eqref{eq:ansatz_def}. The number of quantum circuits required to run VQOS can be calculated in advance from the Hamiltonian and number of variational parameters. The number of quantum circuits required to estimate Eq.~\eqref{eq:N_imp} is $p(p+1)/2$, and for Eq.~\eqref{eq:W_imp}, it is $p n_H$.

\subsection{Effect of estimation errors}
In the implementation above, the parameter velocity is obtained by estimating the matrix $N$ and vector $W$ on a quantum device and then solving the classical linear equation
\begin{equation}
    N\dot{\bm\theta}=W .
\end{equation}
In an ideal noiseless setting, $N$ and $W$ are evaluated exactly. 
In practical implementations, however, their entries are estimated from a finite number of measurements and are affected by imperfections, such as gate and readout errors. 
Therefore, the actual linear system solved by the algorithm is generally written as
\begin{equation}
    (N+\delta N)\dot{\bm\theta}_{\mathrm{est}} = W+\delta W ,
\end{equation}
where $\delta N$ and $\delta W$ denote the errors in the estimated matrix and vector, respectively.

Let $\dot{\bm\theta}$ be the ideal solution satisfying $N\dot{\bm\theta}=W$ and 
\begin{equation}
    \delta\dot{\bm\theta} := \dot{\bm\theta}_{\mathrm{est}}-\dot{\bm\theta}.
\end{equation}
Assuming that the perturbations $\delta N$ and $\delta W$ are sufficiently small, we obtain, to the first order,
\begin{equation}
    N\delta\dot{\bm\theta} \simeq \delta W-\delta N\dot{\bm\theta}.
\end{equation}
Hence,
\begin{equation}
    \delta\dot{\bm\theta} \simeq N^{-1}(\delta W-\delta N\dot{\bm\theta}) .
\end{equation}
This expression shows that the errors in the estimates of $N$ and $W$ are amplified by the inverse of $N$. 
In particular, when $N$ is ill-conditioned, even small statistical or systematic errors can cause large errors in the parameter velocity.

This sensitivity is closely related to the structure of the ansatz. 
The matrix $N$ describes the manner by which changes in the variational parameters translate into changes in the unitary operator $U(\bm\theta)$. 
If two parameters affect $U(\bm\theta)$ almost similarly, then the corresponding parameter directions cannot be distinguished easily. 
Consequently, $N$ has small singular values and the solution of $N\dot{\bm\theta}=W$ becomes sensitive to small errors in the estimated $N$ and $W$. 
Such a situation can occur when the ansatz contains redundant parameters or is overparameterized for the target time-evolution operator.
In such cases, the variational update becomes sensitive to estimation errors. 
Thus, the condition number $N$ is an important quantity that controls the practical stability of VQOS.

For finite-shot measurements, the statistical error of each Pauli expectation value typically scales as $O\left({1}/{\sqrt{N_{\mathrm{shot}}}}\right)$,
where $N_{\mathrm{shot}}$ is the number of measured shots. 
These statistical fluctuations propagate to the entries of $N$ and $W$ and then to $\dot{\bm\theta}$ through the linear solution. 
Moreover, because the parameters are obtained by integrating $\dot{\bm\theta}$ over time, the resulting errors can accumulate during the simulation. 

The gate and readout errors exhibit characteristics different from those of finite sampling noise. 
Although finite-shot noise can be reduced by increasing $N_{\rm shot}$, gate and readout errors generally introduce systematic biases into the estimated values of $N$ and $W$. 
Consequently, the algorithm effectively solves a perturbed variational equation determined by noisy circuits rather than the ideal equation. 
However, because VQOS is designed to approximate time-evolution operators with circuits shallower than those used in Trotterization, the reduced circuit depth may be beneficial for suppressing the accumulation of gate errors.
	
\section{numerical simulation}\label{sec:sim}
In this section, we describe numerical calculations using VQOS to obtain time evolution operators.
\begin{figure}[t]
    \includegraphics[width=8cm]{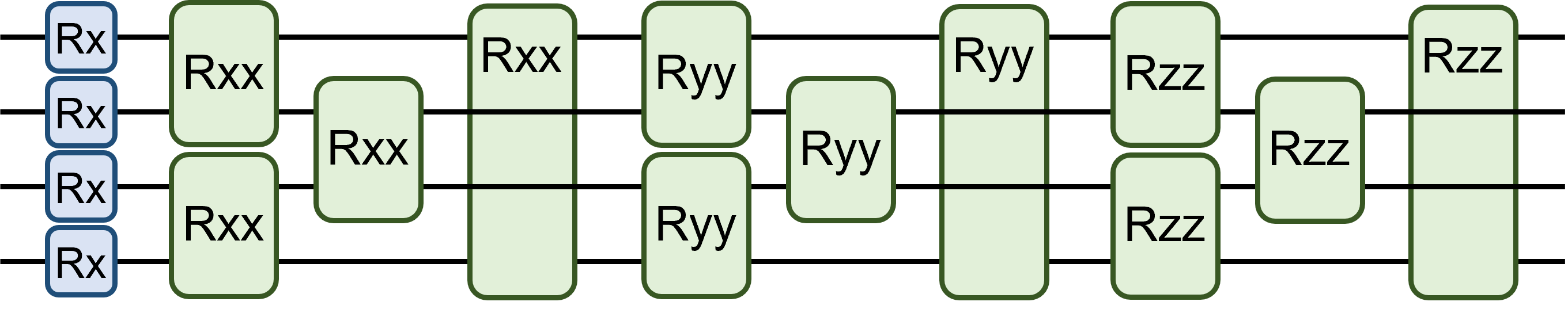}
    \caption{\label{fig:hba}One layer of variational circuit $U(\bm\theta)$ when $n=4$, consisting of Rx gates, Rxx gates, Ryy gates, and Rzz gates, with the same coupling as the Hamiltonian. The Rx gate is defined as $\exp(-i X \theta/2)$ and Rxx gate as $\exp(-i X \otimes X \theta/2)$, and the same for Ryy and Rzz gates. Each rotation gate has different parameters.}
\end{figure}
We adopt the 1D periodic boundary transverse magnetic field Heisenberg model as the system of interest. The Hamiltonian $ H $ is written as follows:
\begin{align}
    H  &= -\frac{1}{2} \sum_{j=1}^nX_j - \frac{1}{2} \sum_{j=1}^n (X_jX_{j+1} + Y_jY_{j+1} + Z_jZ_{j+1}) \nonumber\\
    &=  H _{\mathrm{x}} +  H _{\mathrm{xx}} +  H _{\mathrm{yy}} +  H _{\mathrm{zz}}.
\end{align}
Here, $n$ is the number of sites, and the $(n+1)$th site is considered the first site. The ansatz $U(\bm\theta)$ (Fig.~\ref{fig:hba}) has a one-to-one correspondence of sites to qubits and a circuit with the same configuration as the first-order Trotterization $U_{\mathrm{trot}}(t, L)$, where
\begin{equation}
    U_{\mathrm{trot}}(t, L) = \left( e^{-i H _{\mathrm{x}}t/L} e^{-i H _{\mathrm{xx}}t/L} e^{-i H _{\mathrm{yy}}t/L} e^{-i H _{\mathrm{zz}}t/L} \right)^L
    \label{eq:trotterization}
\end{equation}
is used to compare the depths of the circuits.  In Eq.~\eqref{eq:trotterization}, $t$ and $L$ denote time and number of layers, respectively.
Here, $L$ denotes the fixed number of layers in the variational ansatz and is not a time-step index. 
For each $L$ selected, only the parameter $\bm\theta(t)$ evolves according to Eq.~\eqref{eq:VQOS_pure_param}, while the circuit structure remains fixed. The initial parameters are set to $\bm\theta(0)=0$, such that $U(\bm\theta(0))=I$.
The parameters $\bm\theta$ of the rotation gates in the ansatz $U(\bm\theta)$ are different for each gate. In this study, we performed simulations using classical computers with the \texttt{Yao.jl} package~\cite{yao}. Details about the numerical simulation are described in Appendix \ref{sec:apd_sims}.

\subsection{Noiseless simulation}
\begin{figure}[t]
    \includegraphics[width=8cm]{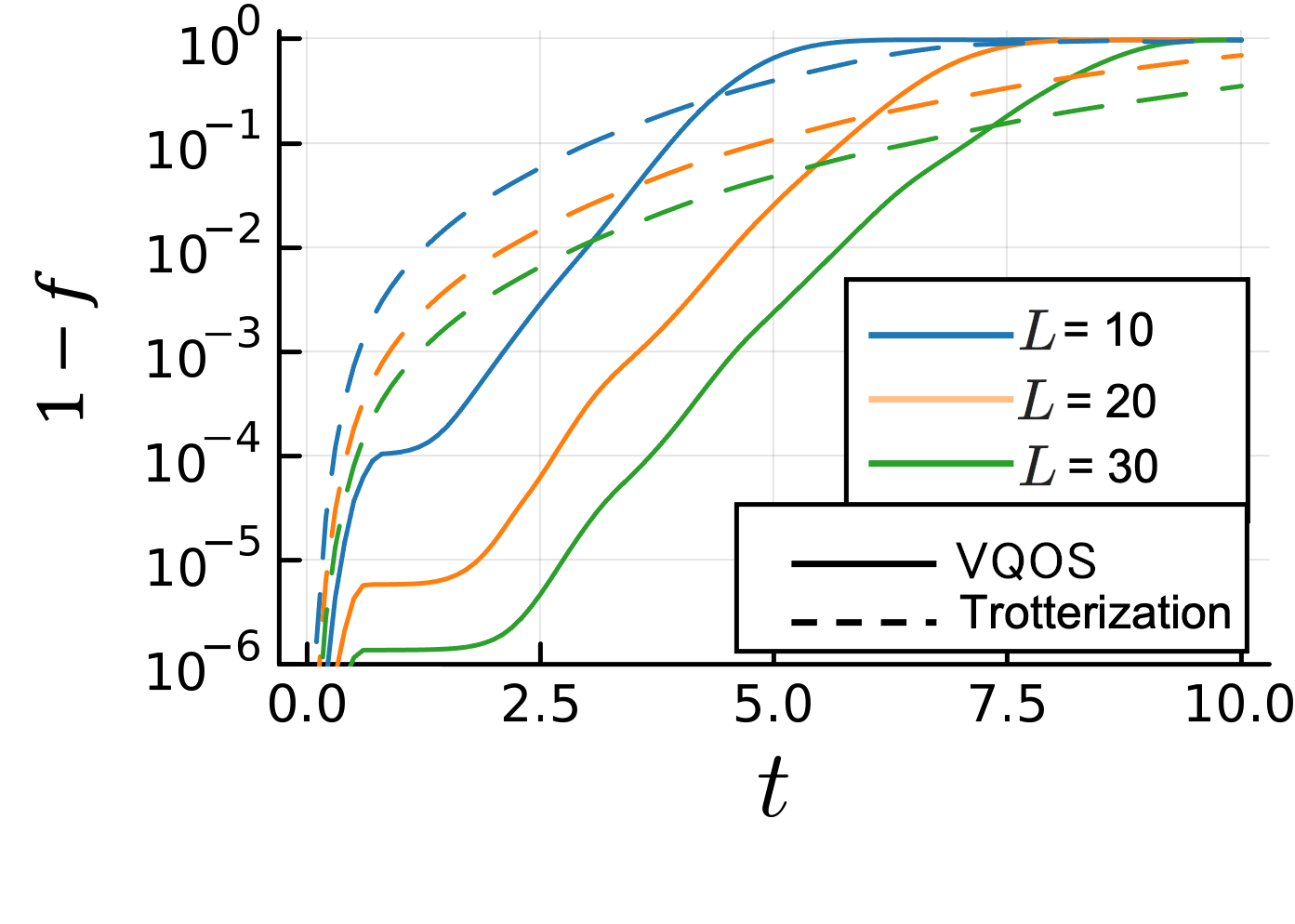}
    \caption{\label{fig:infids_np6-9} The process infidelity $1-f$ of the time evolution operators of the $9$-site 1D periodic boundary transverse magnetic field Heisenberg model obtained by VQOS and Trotterization against the exact operator obtained by diagonalization of the Hamiltonian; the solid line is the result of VQOS, and the dotted line is the result of Trotterization. The blue, orange, and green plots correspond to 10, 20, and 30 layers, respectively. 
    }
\end{figure}

In this subsection, we present the noiseless simulation of VQOS.
Fig.~\ref{fig:infids_np6-9} shows the error of the time evolution operator obtained by VQOS and Trotterization compared to the exact time evolution operator obtained by diagonalization of the Hamiltonian. The error of the two $n$-qubit operators $U$ and $V$ was evaluated by the process infidelity~\cite{operator_fidelity}, which is defined as $1-f = 1 - |\mathrm{Tr}(V^{\dag}U)/2^n|$. We found that in the range where the process infidelity is less than $10^{-1}$, the VQOS error is always less than that of the Trotterization. We also found that the VQOS error increases exponentially with respect to $t$. This is thought to be because the increase in error due to sequential simulations is proportional to the current error. It can also be assumed that the ansatz has sufficient expressibility at the plateau that appears at the beginning of the simulation. Meanwhile, the error of Trotterization increases in proportion to the cube of $t$. This is because the leading term of the first-order Trotterization error is $\mathcal{O}(t^3)$. VQOS accuracy is worse than that of Trotterization in regions where operator infidelity is close to $1$. This indicates that the variational principle simulation does not work well and does not find the optimal parameters within the range allowed by the trial function.

\begin{figure}[t]
    \includegraphics[width=8cm]{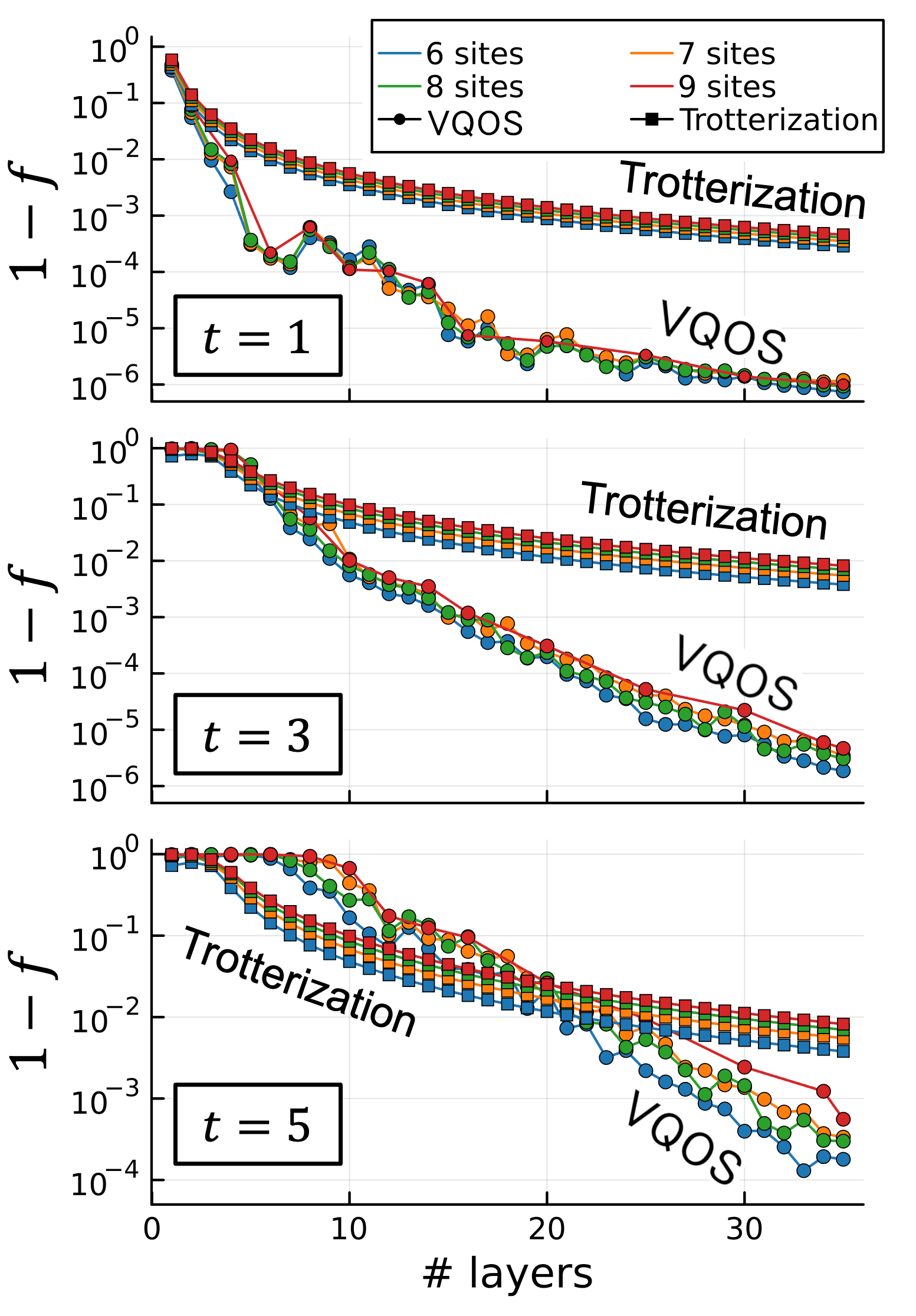}
    \caption{\label{fig:op_infids_lays_VQOS_trot} Relationship between number of layers and operator infidelity $1-f$ at $t=1$, $3$, and $5$ in VQOS and Trotterization. Results for different sites correspond to different colored plots. The plots with round and square markers correspond to the results of VQOS and Trotterization, respectively.}
\end{figure}

Next, Fig.~\ref{fig:op_infids_lays_VQOS_trot} shows the relationship between the number of layers and accuracy of VQOS and Trotterization. VQOS shows better accuracy than Trotterization except when $t$ is large and the number of layers is small. When the number of layers is 30, the accuracy improves by a factor of 10 to 100. Furthermore, VQOS shows an exponential improvement in accuracy with increasing number of layers within the tested parameter range. In addition, when the number of sites is greater than 6, the difference in results among the number of sites is small. The difference in results between different numbers of sites increases as $t$ increases, but the difference is not large at $t=5$. Therefore, the number of layers required for VQOS is expected to scale slowly with increasing number of sites.

\begin{figure}[t]
    \includegraphics[width=8cm]{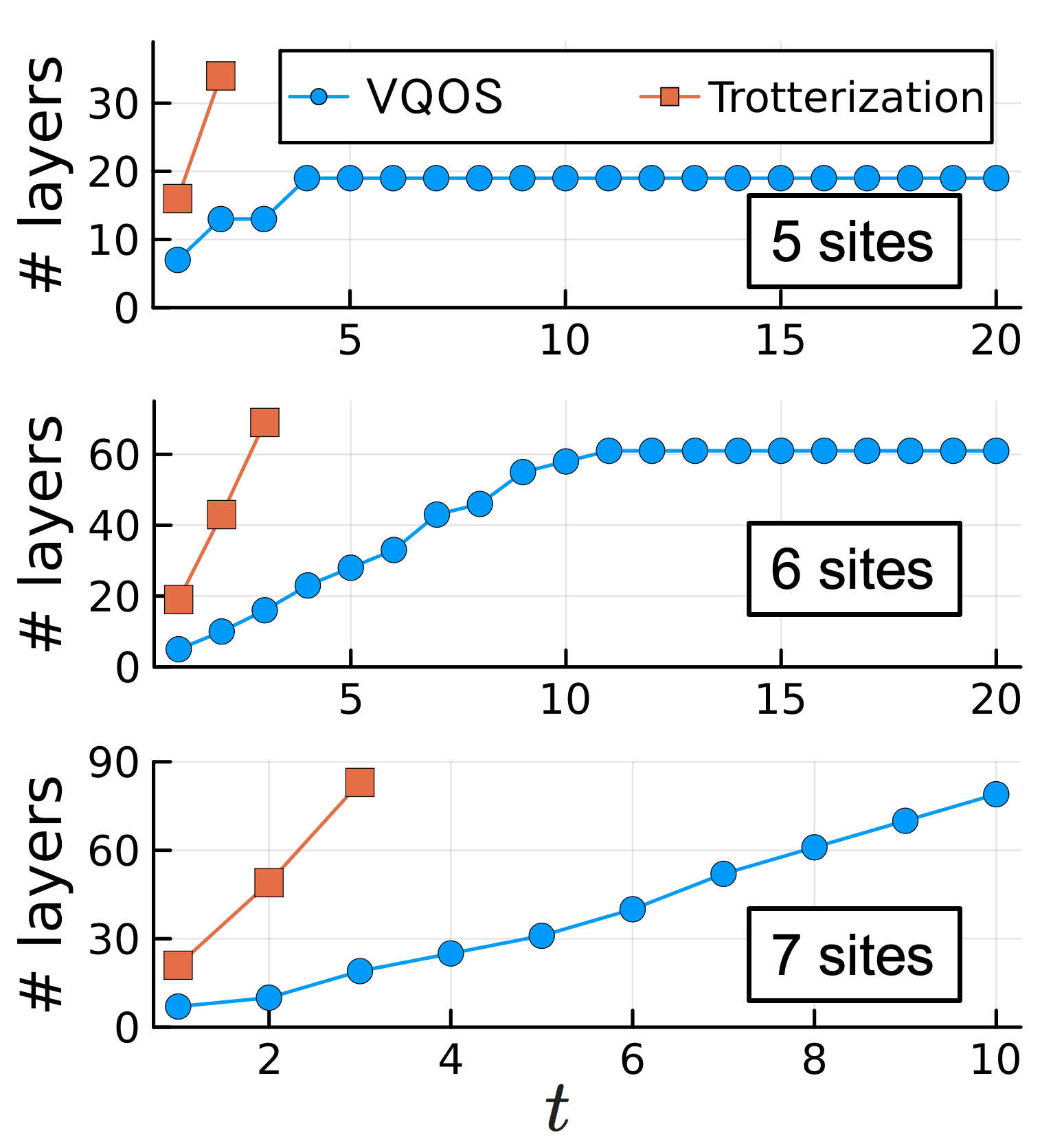}
    \caption{\label{fig:req_layers_VQOS_trot_np5-7}Number of layers required to make operator infidelity $1-f$ smaller than $10^{-3}$ in VQOS and Trotterization for transverse field Heisenberg model with 5--7 sites. The plots with round and square markers correspond to the
        results of VQOS and Trotterization, respectively.}
\end{figure}
Finally, we discuss the number of layers required for VQOS. Fig.~\ref{fig:req_layers_VQOS_trot_np5-7} shows the number of layers required to achieve process infidelity smaller than $10^{-3}$ in VQOS and Trotterization. The number of layers required in VQOS increases linearly with $t$ at first and then saturates. The number of saturating layers increases rapidly with the number of sites, and we could not find saturation in our calculations for more than 7 sites. In the saturating number of layers, the variational quantum circuit appears to have the ability to represent the time evolution operator in general time $t$. As the number of sites increases, the complexity of the general time evolution operator increases, and the number of saturating layers also increases. In any case, the number of layers required for VQOS is much less than for Trotterization.

Additional benchmarks are presented in the appendices. In Appendix~\ref{sec:second_order_trotter}, we compare VQOS with second-order Trotterization by matching the number of two-qubit gates, and find the same qualitative tendency in the low-infidelity regime.  In Appendix~\ref{sec:obc}, we also examine the corresponding model with open boundary conditions and confirm that the qualitative behavior remains unchanged.

\subsection{Finite-shot simulation}
Finally, we examined the effect of finite sampling noise on VQOS. 
As discussed in Sec.~\ref{sec:imp}, the parameter velocity in VQOS is obtained by estimating the matrix $N$ and vector $W$ and then solving the linear equation $N\dot{\bm\theta}=W$. 
Therefore, the statistical errors in the estimates of $N$ and $W$ can be amplified using a linear solution and accumulated during the time integration of the variational parameters.

To investigate this effect, we performed a finite-shot numerical simulation of the three-qubit transverse-field Heisenberg model using an $L=4$ ansatz. 
In this simulation, we assumed that the implementation of VQOS presented in Fig.~\ref{fig:VQS_impl}(a) and the expectation values used to construct $N$ and \(W\) were estimated using a finite number of shots. 
We considered $N_{\mathrm{shot}}=10^3$, $10^4$, and $10^5$ and compared the results with those of the noiseless case. 
Gate noise and readout noise were excluded; therefore, the only source of error in this simulation was the finite sampling noise.

\begin{figure}[t]
    \centering
    \includegraphics[width=9cm]{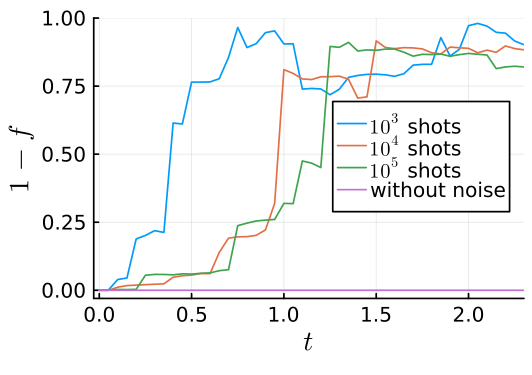}
    \caption{
    Process infidelity $1-f$ of VQOS under finite sampling noise for three-qubit transverse-field Heisenberg model with four-layer ansatz. 
    Results are shown for $N_{\mathrm{shot}}=10^3$, $10^4$, and $10^5$, together with results of noiseless case. 
    Gate noise and readout noise are excluded.
    }
    \label{fig:vqos_finite_shots}
\end{figure}

Fig.~\ref{fig:vqos_finite_shots} shows that finite sampling noise can significantly degrade the accuracy of VQOS. 
Even though the results from the noiseless case were accurate in the plotted time range, the finite-shot results exhibited a rapid increase in the process infidelity. 
This indicates that statistical fluctuations in $N$ and $W$ can significantly affect the parameter updates.
Increasing the number of shots improved the results, as expected from the statistical scaling of measurement errors. 
However, this improvement is insufficient to completely reproduce the noiseless-case results in this example. 
This suggests that practical implementations of VQOS require not only a sufficiently large number of shots but also optimized shot allocation and error mitigation to reduce the effect of statistical errors.
These finite-shot results do not change the noiseless advantage of VQOS demonstrated above; however, they clarify an important practical issue for near-term implementations. 
A more systematic study of the finite-shot effects and realistic device noise shall be endeavored in the future.
	
\section{Conclusions}\label{sec:conc}
In this study, we developed VQOS, a method to approximate and implement the time evolution operator of a quantum system. We emphasize that VQOS is particularly suited for near-term applications in which (i) short- to intermediate-time evolution operators are required and (ii) explicit access to the time-evolution operator itself is essential, such as in quantum phase estimation and quantum eigenvalue transformation.
The method was derived by applying McLachlan's variational principle to a parameterized operator. It does not require another method to implement time evolution operators, such as Trotterization. Furthermore, it is not based on an iterative optimization of the cost functions but can be performed with a predetermined number of quantum operations. In addition, the circuit implementation of VQOS is resource-efficient. The algorithm can be executed without Bell pairs or additional registers, resulting in shallower circuits than a naive Choi-state implementation.

Numerical results using VQOS to approximate the time evolution operator showed an accuracy improvement of up to 3 orders of magnitude or more for the same circuit depth compared to Trotterization, reducing the circuit depth by approximately $1/5$ for the same accuracy comparison. In addition, it is suggested that the ansatz depth of VQOS can scale well with respect to the size of the system.
We also discussed the effect of noise on VQOS. 
In particular, finite sampling errors can be amplified through the linear solve, indicating that shot allocation and noise mitigation will be important for practical implementations.

The method described in the main text of this paper can be extended to the dynamics of open quantum systems governed by the Lindblad equation. The corresponding formulation and implementation are presented in Appendix \ref{sec:apd_open}.

\appendix
\def\thesection{\Alph{section}}
  \section{Derivation of Eq.~\eqref{eq:VQOS_pure_param}}
In this appendix, we derive Eqs.~\eqref{eq:VQOS_pure_param},
~\eqref{eq:N_pure} and \eqref{eq:W_pure}
beginning from the variational principle expressed in Eq.~\eqref{eq:McLachlanVP_OVQS}.
We consider a parameterized unitary
$\tilde{U}(\theta_0(t), \bm\theta(t))=e^{i\theta_0(t)}U(\bm\theta(t))$,
where the size of $\tilde{U}$ is $2^n\times 2^n$;
where $\theta_0(t)$ denotes the global phase.
The time derivative of $\tilde{U}$ is expressed as
\begin{align}
\frac{d\tilde{U}(\theta_0(t), \bm\theta(t))}{dt}
&= \frac{\partial \tilde{U}(\theta_0, \bm\theta)}{\partial \theta_0}
 \dot{\theta}_0
+\sum_j \frac{\partial \tilde{U}(\theta_0, \bm\theta)}{\partial \theta_j}
 \dot{\theta}_j \nonumber \\
&=ie^{i\theta_0}U(\bm\theta) \dot{\theta}_0
+ e^{i\theta_0}
\sum_j \frac{\partial U(\bm\theta)}{\partial \theta_j} \dot{\theta}_j.
\end{align}
The square of the Frobenius norm used for variational analysis is defined as
\begin{equation}
F(\dot{\theta}_0(t), \dot{\bm\theta}(t)) :=
\left\|
 \frac{d\tilde{U}(\theta_0(t), \bm\theta(t)))}{dt}
 + iH\tilde{U}(\theta_0(t), \bm\theta(t))
\right\|^2_{\mathrm{F}},
\end{equation}
where $\|A\|^2_{\mathrm{F}}=\Tr(A^\dagger A)$.
The square of the Frobenius norm is calculated as follows:
\begin{align}
&F(\dot{\theta}_0(t), \dot{\bm\theta}(t))\nonumber \\
=&
2^n\dot{\theta_0}^2
+ \sum_{j k}\Tr\left[
\left(\frac{\partial U(\bm\theta)}{\partial \theta_j}\right)^\dagger
\left(\frac{\partial U(\bm\theta)}{\partial \theta_k}\right)
\right]\dot{\theta_j}\dot{\theta_k}
+ \Tr\left[H^2\right] \nonumber \\
&-2\sum_j\Im\Tr\left[
\left(\frac{\partial U(\bm\theta)}{\partial \theta_j}\right)^\dagger
U(\bm\theta)
\right]\dot{\theta_0}\dot{\theta_j} \nonumber \\
&-2\sum_j\Im\Tr\left[
\left(\frac{\partial U(\bm\theta)}{\partial \theta_j}\right)^\dagger
HU(\bm\theta)
\right]\dot{\theta_j}
+2\Tr[H]\dot{\theta_0}.
\end{align}
Considering infinitesimal changes in $\dot{\theta}_0(t)$,
the condition under which the Frobenius norm attains a stationary value
is as follows:
\begin{equation}
2^n\dot{\theta_0}
 - \sum_j\Im\Tr\left[
\left(\frac{\partial U(\bm\theta)}{\partial \theta_j}\right)^\dagger
U(\bm\theta)\right]\dot{\theta_j}
+ \Tr[H]=0.\label{eq:vari_theta_0}
\end{equation}
The condition under which the Frobenius norm attains a stationary value
when $\dot{\theta}_j(t)$ is varied infinitesimally for a specified $j$ is 
\begin{align}
&\sum_k\Re\Tr\left[
\left(\frac{\partial U(\bm\theta)}{\partial \theta_j}\right)^\dagger
\left(\frac{\partial U(\bm\theta)}{\partial \theta_k}\right)
\right]\dot{\theta_k} \nonumber \\
&-\Im\Tr\left[
\left(\frac{\partial U(\bm\theta)}{\partial \theta_j}\right)^\dagger
U(\bm\theta)\right]\dot{\theta}_0 \nonumber \\
&-\Im\Tr\left[
\left(\frac{\partial U(\bm\theta)}{\partial \theta_j}\right)^\dagger
HU(\bm\theta)\right]=0.\label{eq:vari_theta_j}
\end{align}
Using Eqs.~\eqref{eq:vari_theta_0} and \eqref{eq:vari_theta_j},
we can eliminate $\dot{\theta}_0$
to derive the following equation for $\dot{\bm\theta}$.
\begin{equation}
N \dot{\bm\theta} = W, \label{eq:apA_N_theta_dot_W}
\end{equation}
where
\begin{align}
N_{jk} &= \Re\Tr\left[ \left(
\frac{\partial U({\bm\theta})}{\partial\theta_j}\right)^\dagger
\left(\frac{\partial U({\bm\theta})}
{\partial\theta_k} \right)\right] \nonumber \\
&- \frac{1}{2^n}
\left\{
\Im\Tr\left[\left(
\frac{\partial U({\bm\theta})}{\partial\theta_j}\right)^\dagger
U({\bm\theta})\right]
\right.\nonumber \\
&\left. \hspace*{1cm}\times
\Im\Tr\left[\left(
\frac{\partial U({\bm\theta})}{\partial \theta_k}\right)^\dagger
U({\bm\theta})\right]\right\} \nonumber \\
&= \Re\Tr\left[ \left(
\frac{\partial U({\bm\theta})}{\partial\theta_j}\right)^\dagger
\left(\frac{\partial U({\bm\theta})}
{\partial\theta_k} \right)\right] \nonumber \\
&+ \frac{1}{2^n}
\left\{
\Tr\left[\left(
\frac{\partial U({\bm\theta})}{\partial\theta_j}\right)^\dagger
U({\bm\theta})\right]
\right.\nonumber \\
&\left. \hspace*{1cm}\times
\Tr\left[\left(
\frac{\partial U({\bm\theta})}{\partial \theta_k}\right)^\dagger
U({\bm\theta})\right]\right\}, \label{eq:apA_N_jk}\\
W_j &= \Im\Tr\left[ \left(\fracpd{U({\bm\theta})}{\theta_j}\right)^\dagger
HU({\bm\theta})\right]\nonumber \\
&- \frac{1}{2^n}
\Im\Tr\left[\left(
\frac{\partial U({\bm\theta})}{\partial\theta_j}\right)^\dagger
U({\bm\theta}) \right] \Tr[H]\nonumber \\
&= \Im\Tr\left[ \left(\fracpd{U({\bm\theta})}{\theta_j}\right)^\dagger
HU({\bm\theta})\right]\nonumber \\
&+\frac{i}{2^n}
\Tr\left[\left(
\frac{\partial U({\bm\theta})}{\partial\theta_j}\right)^\dagger
U({\bm\theta}) \right] \Tr[H]. \label{eq:apA_W_j}
\end{align}
Here, we consider the fact that
$\Tr\left[\left(
\frac{\partial U({\bm\theta})}{\partial\theta_j}\right)^\dagger
U({\bm\theta}) \right]$
is a purely imaginary number
that is obtained by taking the partial derivative of
$U^\dagger({\bm\theta})U({\bm\theta}) = I$
with respect to $\theta_j$.
Although $\theta_0(t)$ does not appear explicitly
in Eqs.~\eqref{eq:apA_N_theta_dot_W} , \eqref{eq:apA_N_jk}, and
\eqref{eq:apA_W_j}, 
the global phase is optimized via variational optimization
using equation \eqref{eq:vari_theta_0}.

\section{Details of numerical simulations\label{sec:apd_sims}}
In this study, we simulated VQOS on a classical computer. We used a method optimized for classical simulators that differs from the method mentioned in Sec.~\ref{sec:imp}. That method reduces redundant computations by first obtaining the derivative of the ansatz $\fracpd{\ket{\phi(\bm\theta)}}{\theta_j}$ for each parameter $\theta_j \in \bm\theta$, and then $\fracpd{\bra{\phi(\bm\theta)}}{\theta_j} \fracpd{\ket{\phi(\bm\theta)}}{\theta_j}$ and $\fracpd{\bra{\phi(\bm\theta)}}{\theta_j}  H  \ket{{\phi(\bm\theta)}}$ are calculated. In this case, the computational cost to obtain $N$ and $W$ in Eq.~\eqref{eq:VQOS_pure_param} is linear with respect to the number of parameters $L$, while that of the method presented in the main text is $\mathcal{O}(L^2)$. This method requires the entire state vector to be stored and thus cannot be used in a real quantum computer.

Following ~\cite{AVQDS}, we added small values $10^{-8}$ to the diagonal components of $N$ before solving Eq.~\eqref{eq:VQOS_pure_param} to avoid the numerical instability that comes with ill-conditioned $N$. The time step of integrating $\dot{\bm\theta}$ to obtain the time evolution of $\bm\theta$ was fixed at $0.05$, and a fourth-order Runge-Kutta method was used.

\section{Comparison with second-order Trotterization}
\label{sec:second_order_trotter}
In the main text, we compared VQOS with first-order Trotterization using circuits with the same number of layers. As an additional benchmark, we compared VQOS with second-order Trotterization to examine whether the qualitative advantage of VQOS remained when a higher-order product formula was used as the reference method. Here, we used the same ansatz as that in the main text for VQOS.

However, this comparison is not entirely equivalent to that presented in Fig.~\ref{fig:infids_np6-9}. In the main-text comparison, the ansatz used for VQOS has the same layer structure as the first-order Trotterization circuit; therefore, the number of layers provides a natural measure of the circuit depth. By contrast, the circuit structure of the second-order Trotterization is different from that of the VQOS ansatz used here. Therefore, instead of matching the number of layers, we compared the two methods by matching the number of two-qubit gates, denoted by $N_{2Q}$. Although this provides a useful resource-based comparison, it should be regarded as an approximate benchmark instead of a completely identical comparison.

\begin{figure}[t]
    \centering
    \includegraphics[width=8cm]{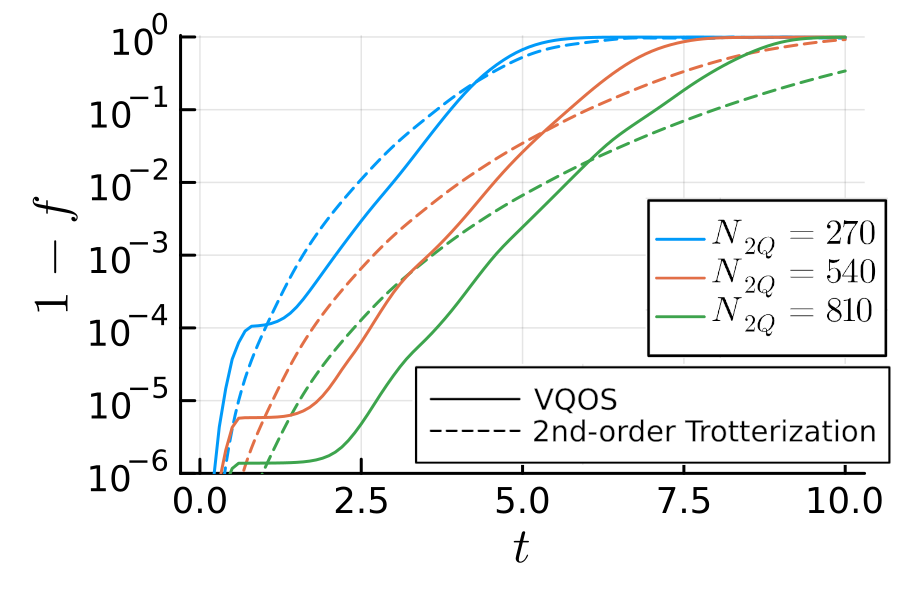}
    \caption{
    Process infidelity $1-f$ of time-evolution operators obtained via VQOS and second-order Trotterization for nine-site transverse-field Heisenberg model. Solid and dashed lines show VQOS and second-order Trotterization results, respectively. Blue, orange, and green curves correspond to $N_{2Q}=270$, $540$, and $810$, respectively, where $N_{2Q}$ denotes number of two-qubit gates. Since the circuit structures of the VQOS ansatz and second-order Trotterization are different, the comparison is performed by matching $N_{2Q}$ instead of by the number of layers.
    }
    \label{fig:vqos_second_order_trotter}
\end{figure}

Fig.~\ref{fig:vqos_second_order_trotter} illustrates the process
infidelity $1-f$ for a nine-site transverse-field Heisenberg model.
This figure is analogous to Fig.~\ref{fig:infids_np6-9} in the
main text, with the exception that the dashed lines represent the results
obtained via second-order Trotterization. The solid lines show
the VQOS results, whereas the blue, orange, and green curves
correspond to the cases with the same number of two-qubit
gates and $N_{2Q}=270$, $540$, and $810$, respectively.
We observed that the qualitative trend observed in the main text remained unchanged. In particular, in the region where the process infidelity was sufficiently small, VQOS achieved a smaller error than the second-order Trotterization for the same number of two-qubit gates. At longer times, i.e., when the infidelity approached $1$, VQOS no longer provided an accurate approximation of the target time-evolution operator, and the advantage of VQOS was lost. This is consistent with the behavior observed in the comparison with first-order Trotterization.

\section{Numerical results under open boundary conditions}\label{sec:obc}

In the main text, we presented the numerical results for a 1D transverse-field Heisenberg model with periodic-boundary conditions. 
To examine whether the qualitative behavior of VQOS depends on the boundary condition, we additionally performed numerical simulations for the corresponding model with open-boundary conditions (OBCs). 
The Hamiltonian is expressed as
\begin{equation}
H_{\mathrm{OBC}} = -\frac{1}{2}\sum_{j=1}^{n} X_j -\frac{1}{2}\sum_{j=1}^{n-1}\left(X_j X_{j+1} + Y_j Y_{j+1} + Z_j Z_{j+1}\right).
\end{equation}
The ansatz was selected in the same manner as in the periodic-boundary case, namely, a layered circuit composed of Rx, Rxx, Ryy, and Rzz gates, following the connectivity of the target Hamiltonian but without boundary coupling between the first and last sites.

\begin{figure}[t]
    \centering
    \includegraphics[width=8cm]{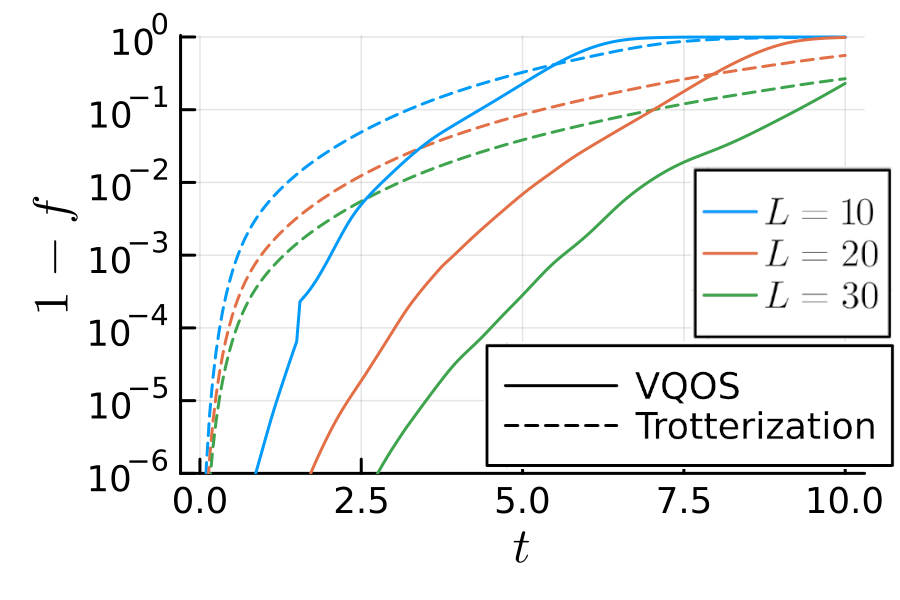}
    \caption{
    Process infidelity $1-f$ of time-evolution operators of nine-site transverse-field Heisenberg model with open-boundary conditions, obtained via VQOS and Trotterization, against exact operator obtained by diagonalizing the Hamiltonian. 
    Solid and dashed lines show VQOS and Trotterization results, respectively. 
    Blue, orange, and green plots correspond to $L=10$, $20$, and $30$ layers, respectively. 
    Qualitative behavior is consistent with that in periodic-boundary case shown in Fig.~\ref{fig:infids_np6-9} of the main text.
    }
    \label{fig:obc_infidelity}
\end{figure}

Fig.~\ref{fig:obc_infidelity} shows the process infidelity of the time-evolution operator obtained via VQOS and first-order Trotterization for the nine-site OBC model. 
The meaning of this figure is the same as that of Fig.~\ref{fig:infids_np6-9} in the main text: the solid and dashed lines represent the VQOS and Trotterization results, respectively; whereas the blue, orange, and green curves correspond to $L=10$, $20$ and $30$, respectively.
We observed that the qualitative trend remained the same as that of the periodic-boundary case. 
In particular, in the region where the process infidelity was sufficiently low, VQOS achieved a smaller error than Trotterization at the same circuit depth. 
As in the periodic-boundary case, the accuracy improved as the number of layers increased, whereas both methods eventually exhibited reduced accuracy at longer times. 
Therefore, the advantage of VQOS over Trotterization observed in the main text is not specific to periodic-boundary conditions but persists under OBCs.

\section{Extension to open dynamics\label{sec:apd_open}}
VQOS can be extended to open quantum systems to find a channel $\mathcal{E} (\bm\theta (t))$ that approximates the time evolution of open systems based on the Lindblad equation, where $\mathcal{E} (\bm\theta (t))$ is the quantum channel parameterized by time-dependent parameters $\bm\theta(t)$. We call this method Variational Quantum Channel Simulation (VQCS). For this purpose, we use the following variational principle:
\begin{equation}
    \delta \left\| \frac{d}{dt} \mathcal{E}(\bm\theta(t)) - \mathcal{L} \circ \mathcal{E} (\bm\theta(t)) \right\| = 0.
    \label{eq:variation_lindblad}
\end{equation}
Here, we define the norm of a super-operator by
\begin{equation}
    \left\| \mathcal{E} \right\| := \sum_a \Tr\left[(\mathcal{E} (X_a))^\dag \mathcal{E} (X_a)) \right],
    \label{eq:def_channel_norm}
\end{equation}
using the inner product of super-operators as defined in \cite{channel_inner_prod}, where $\{{X_a}\}$ is an orthonormal basis of input systems of $\mathcal{E}$. Using Eq.~\eqref{eq:def_channel_norm}, the variational principle of Eq.~\eqref{eq:variation_lindblad} gives the equation of parameters for open systems
\begin{equation}
    N \dot{\bm\theta} = W,
    \label{eq:VQCS_density_NW}
\end{equation}
where
\begin{align}
    N_{jk} &= \Tr \left[ \left( \fracpd{\mathcal{E}(\bm\theta)}{\theta_j} \otimes \mathcal{I} \right) (\ketbra{\Phi}{\Phi})\nonumber \right. \\ 
    &\hspace{8em} \left. \left( \fracpd{\mathcal{E}(\bm\theta)}{\theta_k} \otimes \mathcal{I} \right) (\ketbra{\Phi}{\Phi}) \right],\\
    W_j &= \Tr \left[ \left( \fracpd{\mathcal{E}(\bm\theta)}{\theta_j} \otimes \mathcal{I} \right) (\ketbra{\Phi}{\Phi}) \mathcal{L} \circ \mathcal{E}(\bm\theta) (\ketbra{\Phi}{\Phi}) \right].
\end{align}
$\mathcal{I}$ is the identity channel.
As in the closed system case, VQCS for open systems is also equivalent to the time evolution of the Choi state of $\mathcal{E}(\bm\theta(t))$ by Lindbladian $\mathcal{L}\otimes\mathcal{I}$ in VQSS for open systems.

\begin{figure*}[t]
    \centering
    \includegraphics[width=14cm]{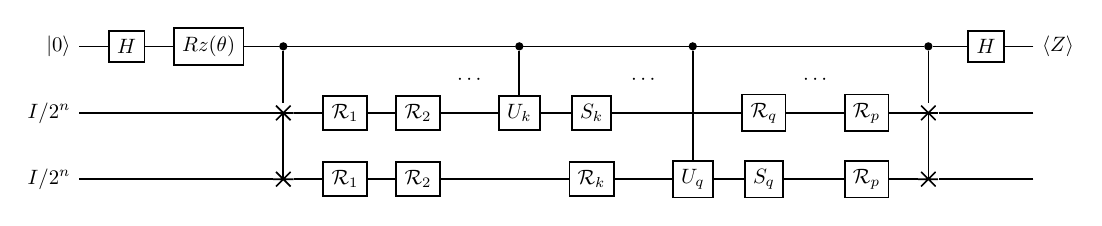}
    \caption{\label{fig:VQCS_impl} Circuit for the evaluation of \eqref{eq:VQCS_impl_g} based on the circuit proposed in \cite{VQST}. The ancillary qubit is initialized in the state $\ket{0}$, and both of the $n$ qubit registers are initialized in the maximally mixed state $I/2^n$. Eq.~\eqref{eq:VQCS_impl_g} is evaluated by estimating the expectation value $\ev{Z}$ of the ancillary qubit. Here, $U_k = T_k S_k^\dag$ and $U_q = T_q S_q^\dag$. In the figure, we assume that $k<q$ and $q=p+1$.}
\end{figure*}

Next, we consider the implementation of VQCS for open systems. Because VQCS is equivalent to performing VQSS on the Choi states, we can use the implementation of VQSS proposed in \cite{VQST}. Suppose that the parameterized quantum channel is given by $\mathcal{E}(\bm\theta) = \mathcal{R}_p(\theta_p) \cdots \mathcal{R}_1(\theta_1)$, where $\mathcal{R}_j(\theta_j)$ are quantum channels parameterized by a real parameter $\theta_j$. We can rewrite the generator as
\begin{equation}
    \mathcal{L}(\rho) = \sum_i g_i S_i \rho T_i^\dag,
\end{equation}
where $S_i$ and $T_i$ are unitary operators, and $g_i$ are coefficients. Similarly, we write
\begin{equation}
    \fracpd{\mathcal{R}_k(\rho)}{\theta_k} = \sum_i r_{k,i} S_{k,i} \rho T^\dag_{k,i},
\end{equation}
where $S_{k,i}$ and $T_{k,i}$ are unitary operators, and $r_{k,i}$ are coefficients. $N$ and $W$ in Eq.~\eqref{eq:VQCS_density_NW} can be expressed as
\begin{equation}
    \begin{aligned}
        N_{k,q} &= \Re\left( r_{k,i} r_{q,j} \Tr\left[ ((\mathcal{R}_{k,i} \otimes \mathcal{I}) \ketbra{\Phi}{\Phi}) \right. \right. \\
        &\hspace{6em}\left. \left. ((\mathcal{R}_{q,j} \otimes \mathcal{I}) \ketbra{\Phi}{\Phi}) \right] \right),\\
        W_{k} &= \Re \left( r_{k,i} g_j \Tr \left[ ((\mathcal{R}_{k,i} \otimes \mathcal{I}) \ketbra{\Phi}{\Phi}) \right. \right. \\
        &\hspace{6em} ((S_j \otimes I) \ketbra{\Phi}{\Phi} (T_j^\dag \otimes I) ) ]).
        \label{eq:VQCS_open_nw_impl}
    \end{aligned}
\end{equation}
Eq.~\eqref{eq:VQCS_open_nw_impl} can be evaluated by estimating
\begin{equation}
    \Re \left[ e^{i\theta} \Tr (\rho_1 \rho_2) \right],
    \label{eq:VQCS_impl_g}
\end{equation}
where
\begin{equation}
    \begin{aligned}
        \rho_1 &= (\mathcal{R}_p \cdots \mathcal{R}_{k+1} \otimes \mathcal{I})\\
        &\hspace*{2em} [(S_k \otimes I)((\mathcal{R}_{k-1 } \cdots \mathcal{R}_1 \otimes \mathcal{I}) \ketbra{\Phi}{\Phi})(T_k^\dag \otimes I)],\\
        \rho_2 &= (\mathcal{R}_p \cdots \mathcal{R}_{q+1} \otimes \mathcal{I})\\
        &\hspace*{2em} [(S_q \otimes I)((\mathcal{R}_{q-1 } \cdots \mathcal{R}_1 \otimes \mathcal{I}) \ketbra{\Phi}{\Phi})(T_q^\dag \otimes I)].\\
    \end{aligned}
\end{equation}
As in the closed system case, because no gate operations are performed on half of the qubits, tracing them out results in the circuit shown in Fig.~\ref{fig:VQCS_impl}.

\bibliography{bibs}
	
\end{document}